\documentclass[10pt,sigconf,letterpaper,nonacm]{acmart}
\usepackage{pgfplots}
\usepgfplotslibrary{groupplots,dateplot}
\usepackage{pgfplotstable}
\usepackage{xcolor}
\usepackage{booktabs}
\usepackage{array}

\usepackage{hyperref}
\usepackage{adjustbox}
\usepackage{makecell}
\usepackage{comment}
\usepackage{balance}
\usepackage{pifont}

\usepackage{amssymb}
\usepackage{xurl}
\usepackage[T1]{fontenc}

\newcommand{\eg}{\textit{e.g.,~}}

\newcommand{\etal}{\textit{et~al.}}

\pgfplotsset{compat=1.18}
\usetikzlibrary{pgfplots.dateplot}

\definecolor{lineblue}{HTML}{1F5FA8}
\definecolor{shutdownred}{HTML}{C62828}
\definecolor{shutdowngreen}{HTML}{2E7D32}
\definecolor{anomalypurple}{HTML}{6A1B9A}
\definecolor{artifactamber}{HTML}{E8A200}

\pgfplotsset{
  iranstyle/.style={
    date coordinates in=x,
    xticklabel style={rotate=45, anchor=north east, font=\footnotesize},
    ylabel={Visible hosts (millions)},
    ylabel style={font=\small},
    yticklabel style={font=\footnotesize},
    grid=both,
    grid style={line width=0.3pt, draw=gray!20},
    major grid style={line width=0.4pt, draw=gray!35},
    width=\linewidth,
    axis line style={line width=0.5pt},
    tick style={line width=0.5pt},
    legend style={
      at={(0.02,0.98)}, anchor=north west,
      font=\footnotesize, draw=gray!50,
      fill=white, rounded corners=2pt, row sep=1pt,
    },
    clip=false,
  }
}

\AtBeginDocument{%
  }

\begin{document}

\title[A Multi-Perspective Study of the Internet Shutdown in Iran]{A Multi-Perspective Study of the\\ Internet Shutdown in Iran}

\author{Ali Sadeghi Jahromi}
\email{alisadeghijahromi@cmail.carleton.ca}
\affiliation{
  \institution{Carleton University}
  \city{Ottawa}
  \state{Ontario}
  \country{Canada}
}

\author{Jason Jaskolka}
\email{jason.jaskolka@carleton.ca}
\affiliation{
  \institution{Carleton University}
  \city{Ottawa}
  \state{Ontario}
  \country{Canada}
}

\begin{abstract}
Iran conducted two nationwide Internet shutdowns in 2026, on January 8--25 and March 1--May~26, the latter lasting 86 days. We characterize both using three complementary measurement planes: six months of daily Censys scan data, BGP analysis of RIPE RIS snapshots spanning 2019--2026, and continuous per-prefix TCP probing from five vantage points. Each plane captures a different aspect of Iranian connectivity, and interpreting any one in isolation can be misleading. Unlike the partial BGP withdrawal of 2019, the 2022 and 2026 shutdowns were enforced by forwarding-plane discard while 80–88\% of Iranian prefixes remained announced, leaving control-plane monitors blind. Restoration is similarly invisible to BGP, appearing in our forwarding-plane measurements as a centrally coordinated step. Censys host counts overshoot to approximately $3.6\times$ their pre-shutdown baseline after both restorations, rather than returning to baseline. Active probing reveals this inflation to be an artifact: most of the ${\sim}3$M apparent hosts are injected UDP/5353 responses synthesized by an on-path element at Iran's international gateway. Finally, AS-path classification shows that some apparent shutdown survivors were routed through foreign upstreams and never traversed the enforcement point. Together, these results show that measuring shutdowns requires reading multiple planes against one another, as no single signal reliably distinguishes genuine connectivity from its absence.
\begingroup
\renewcommand{\thefootnote}{}
\footnotetext{This manuscript is a preprint. We welcome feedback and suggestions.}
\endgroup
\end{abstract}

\maketitle

\section{Introduction}
\label{sec:introduction}
Internet censorship, which involves fine-grained traffic manipulation, has been documented in Iran for over a decade~\cite{verkamp2012inferring,aryan2013internet,anderson2013dimming}, with prior work characterizing mechanisms such as protocol filtering and keyword-based blocking~\cite{pearce2017global,pearce2017augur,tai2025irblock,webquickelmenhorst21,censorsihpDNS}. Less attention has been given to large-scale, coarse-grained, event-driven Internet disruptions, which have been observed in multiple countries during periods of unrest or crisis, including Egypt and Libya in 2011~\cite{dainotti2011analysis}, and Myanmar in 2021~\cite{padmanabhan2021multi}.\footnote{\url{https://www.accessnow.org/internet-shutdowns-2025/}} Iran has also experienced such disruptions, including during the 2019 fuel-price protests, the 2022 Mahsa Amini period, and two nationwide events in 2026 that we study in this paper. 

Iran’s Telecommunication Infrastructure Company (TIC, AS49666) operates a centralized international gateway. Rather than withdrawing routes, TIC enforces shutdowns by installing forwarding-plane null routes that silently discard traffic while leaving BGP announcements intact. As a result, Iran remains visible to BGP-based monitors during nationwide outages. Using RIPE RIS snapshots across three events (2019--2026), we show 80--88\% of RIPE-allocated Iranian IPv4 prefixes remain globally announced during the 2022 and 2026 shutdowns, with within-event variation under 1.7 percentage points (pp). In contrast, in 2019, coverage drops from 85.1\% to 62.3\% and 54.7\%, indicating partial BGP withdrawal. This shift suggests an evolution from a mixed control- and forwarding-plane approach in 2019 to pure forwarding-plane null routing in later events.

We analyze two nationwide shutdowns, from January 8--25 and March~1--May~26, the latter lasting 86 days and representing the longest documented Iranian disruption, using a multi-plane approach that combines passive Censys scan data, BGP routing analysis, and active per-prefix reachability probing from five global vantage points. The January event reduces visible hosts by 98\% from a baseline of approximately 935K; the March event is more severe, reaching a floor of approximately 10--11K hosts, about 1\% of the historically validated baseline, of which only a third are actively responding at the deepest point.

Our central observation is that each plane can give a misleading picture of Iranian connectivity if investigated alone, and that the three can mislead in distinct and separable ways. BGP announcements persist through both shutdowns and restoration, leaving control-plane monitors blind to both events. Censys host counts do not return to baseline after either restoration but instead overshoot it by approximately $3.6\times$ and remain elevated. Active probing shows this inflation to be largely an artifact of response injection: an on-path element at Iran's gateway synthesizes replies on UDP port~5353 across announced address space, including network and broadcast addresses that cannot be assigned to hosts, so the $\sim$3M apparent additional hosts are injected, not genuine. Finally, geolocation of announced prefixes overstates in-country survival during shutdowns. AS-path classification shows that a substantial fraction of apparent shutdown survivors, including ArvanCloud CDN, were routed through foreign upstreams and never traversed the enforcement point. Characterizing a shutdown therefore requires combining forwarding- and routing-plane measurements and using their discrepancies to classify enforcement rather than merely detect it. In summary, this paper makes the following contributions:

\begin{itemize}

    \item Using six months of passive Censys scan data and active probing, we show that the $3.6\times$ post-restoration host-count inflation is largely an artifact of response injection: an on-path element at Iran's gateway synthesizes UDP/5353 replies across announced address space, inflating observed counts by millions of phantom hosts. This demonstrates that passive host counts cannot be interpreted as connectivity signals across enforcement transitions and identifies a concrete injection signature that future scan-based studies can detect~(Sections~\ref{sec:censys} and~\ref{sec:census}).

    \item From the same passive measurements, combined with BGP AS-path classification, we distinguish genuine shutdown survivors from geolocation artifacts, showing that the networks reachable behind TIC are operational infrastructure (mobile core, state backbone, and TIC's own gateway) while several prominent apparent survivors were never subject to enforcement~(Section~\ref{sec:as-analysis}).

    \item We document a longitudinal shift in enforcement mechanisms using a consistent methodology across 2019, 2022, and 2026, from partial BGP withdrawal to enforcement entirely below the control plane~(Section~\ref{sec:bgp-routing}).

    \item We present a per-prefix forwarding-plane characterization of a national shutdown and its restoration, probing every announced Iranian prefix from five vantage points, and show that restoration was abrupt and centrally coordinated, while remaining invisible to BGP~(Section~\ref{sec:census}).

\end{itemize}

\section{Background and Related Work}
\label{sec:background}
Prior work has developed complementary approaches to detecting and characterizing Internet shutdowns. Early studies by Dainotti~\etal~\cite{dainotti2011analysis} demonstrated the use of BGP and darknet traffic to analyze the 2011 shutdowns in Egypt and Libya. More recent work by Ramesh~\etal~\cite{ramesh2023network} applied multi-plane measurements to Russia's 2022 network responses, identifying control-plane changes and forwarding-plane filtering. The OONI project~\cite{filasto2012ooni} and Aryan et~al.~\cite{aryan2013internet} have documented Iranian application-layer
blocking through in-country measurement. Aceto~\etal~\cite{aceto2026iran} investigate the January~2026 shutdown using aggregated third-party and transit-traffic data, including circumvention via Starlink and peer-to-peer mesh tools. In contrast, we perform per-prefix active reachability probing across all BGP-visible Iranian prefixes, extend the analysis to the March~2026 shutdown, and show that gateway response injection inflates passive scan counts, causing scan-based studies to misreport shutdown severity.

Internet Outage Detection and Analysis (IODA)~\cite{ioda} integrates BGP visibility, active ICMP probing, and Internet background radiation for near-real-time outage detection. Durumeric~\etal~\cite{durumeric2025censys} provide a detailed description of the Censys Internet scanning infrastructure, including its scanning methodology and data processing pipeline. A comparative IODA analysis of Iranian shutdowns~\cite{ioda-comparative} characterizes the 2019 event as withdrawal-based, the 2022 event as mobile-scoped nightly curfews that produced a small diurnal dip in aggregate routing announcements, and the 2025 and 2026 events as occurring without routing withdrawal. These findings independently corroborate the mechanism shift we quantify at the per-prefix level. Their continuous, country-aggregate signals resolve sub-daily dynamics that our discrete snapshots cannot; our measurements instead resolve which prefixes are affected, distinguish forwarding-plane discard from route withdrawal at prefix granularity across multiple vantage points, and identify gateway response injection that inflates passive host counts.

Our work uses cross-plane discrepancies as the primary indicator of disruption: BGP stability can mask forwarding-plane enforcement, and passive scan counts can be inflated by injection even as reachability is unchanged. Platforms such as IODA compare routing, active-probing, and telescope signals across these same events~\cite{ioda-comparative} and observe the mechanism shift away from routing withdrawal. We extend this view by resolving disruption at prefix granularity, distinguishing forwarding-plane discard from route withdrawal, separating genuine survivors from foreign-routed address space, and identifying gateway response injection that inflates passive scan counts but is missed by aggregate connectivity signals.

\subsection{Iranian Internet Architecture}
\label{sec:overview}
Iran operates a highly centralized Internet architecture in which the Telecommunication Infrastructure Company (TIC) and Institute for Research in Fundamental Sciences (IPM) function as the primary
international transit gateways~\cite{anderson2013dimming,madory2026iranian_shutdown}. Domestic operators rely on TIC for upstream access, creating an enforcement point at the international boundary. This centralized topology enables coordinated, near-simultaneous disruption across independently operated ASes without requiring independent action by individual networks.

In parallel, Iran has developed the National Information Network (NIN)~\cite{aryan2013internet,mehr2016nin_launch}, a domestically scoped infrastructure that maintains internal connectivity independent of global routing. The NIN is the operational foundation for prolonged shutdowns, preserving domestic reachability while allowing TIC to enforce forwarding-plane null routes at the international boundary and keep government and critical services functioning internally. Separately, approximately 12--20\% of RIPE-allocated Iranian IPv4 prefixes are not globally announced under normal conditions, held in reserve or used for address space that does not require international reachability. This property is stable across all measurement dates rather than a shutdown artifact, and motivates our use of RIPE-delegated address space as the denominator for all coverage and severity calculations.
\section{Methodology}
\label{sec:method}
\begin{figure}[ht!]
\centering
\includegraphics[width=0.95\linewidth]{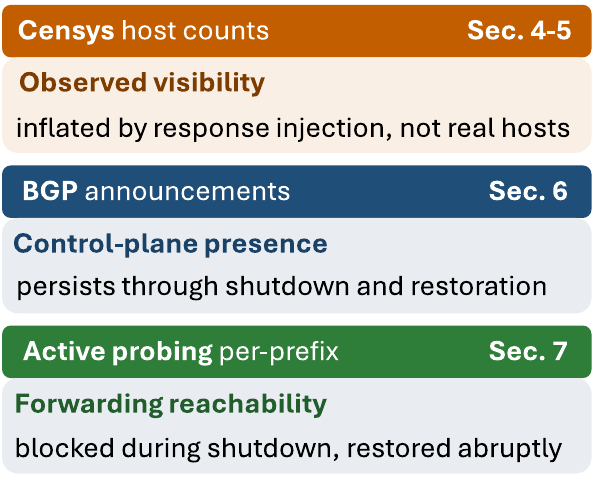}
\caption{The three measurement planes: BGP shows control-plane presence, active probing forwarding-plane reachability, and Censys observed host visibility. Reading them together reveals enforcement that no single signal exposes.} 
\label{fig:summary}
\end{figure}

We combine three measurement planes, passive scanning, active probing, and BGP routing, and treat the discrepancies between them as the primary object of study. When one plane reports connectivity that another contradicts, the disagreement localizes the enforcement mechanism (Figure~\ref{fig:summary}). All three planes measure connectivity to the global Internet; Iran's domestic NIN is outside our vantage and not observed.

\paragraph{Passive measurement (Censys).}
We use the Censys BigQuery Universal Internet Dataset of daily Internet-wide scans~\cite{durumeric2025censys}, querying distinct IPv4 addresses geolocated to Iran for each day from January~1 to June~30, 2026 (178 observations; January 9--11 are absent), together with each host's origin AS and covering \texttt{/24}. Because Censys retains unresponsive hosts for up to 72 hours before removal, we classify hosts as active when none of their services carry a \texttt{pending\_removal\_since} timestamp and as pending eviction otherwise. This field is unpopulated dataset-wide from May~15 onward, as confirmed by our control, so pending-based analysis is restricted to earlier dates. Twelve monthly 2025 snapshots establish a stable baseline of ${\approx}910$K--1.17M hosts, and the January 2026 value (approximately $935$K) serves as our reference denominator. These snapshots are drawn from the 15th of each month. Iran's June 2025 shutdown (approximately June 17–24)~\cite{ioda-comparative} therefore falls between the June~15 and July~15 samples; unlike our daily-resolution 2026 series, this monthly cadence is not designed to resolve week-scale disruptions occurring between sampling dates, and the 2025 baseline should be read accordingly. We use Turkey as a control, querying matched dates in 2026 and 2025; its stable ${\approx}1.3$--$1.5$M host count allows us to attribute Iran-specific variation to the shutdowns.

\paragraph{BGP routing analysis.}
We parse RIPE RIS \texttt{rrc00} BGP snapshots~\cite{ripenis} across the 2019, 2022, and 2026 events, sampling each from baseline through onset, floor, restoration, and the post-restoration plateau, and identify Iranian-origin prefixes using date-correct ASN sets from RIPE NCC delegated statistics. \emph{BGP coverage} is the fraction of RIPE-allocated Iranian IPv4 prefixes with a covering announcement, including exact, more-specific, or aggregate routes. Cross-checks against \texttt{rrc03} and \texttt{rrc04} deviate by at most 1.7 percentage points. To distinguish genuine survivors from address space that only geolocates to Iran, we additionally classify each candidate AS by whether its observed AS paths traverse an Iranian transit AS, including TIC (AS49666/48159) and IPM (AS6736), and report the per-peer fraction crossing Iran (Section~\ref{sec:as-analysis}).

\paragraph{Active probing.}
We deploy agents on five VPSes in Istanbul, Frankfurt, Amsterdam, Singapore, and New York, with opt-out webpages and reverse-DNS identification following responsible scanning practices~\cite{durumeric2024ten}. For each announced Iranian prefix, we probe a fixed host address via TCP on ports 80, 443, and 179 (BGP); an RST on 179 with timeouts on 80/443 is indicative of reachable routing infrastructure, but an all-ports timeout can also arise from an unresponsive host or a firewall dropping unsolicited traffic, so it does not by itself establish forwarding-plane discard. Each prefix receives one of five verdicts (silent\_discard, unreachable, firewall\_acl, reachable, ambiguous) by majority vote across runs. Probing one address per prefix and unable to separate enforcement from firewall silence, these measurements capture reachability breadth as a loose lower bound, not host-level depth; we interpret only aggregate rates and their temporal change, not individual verdicts. We run a dense census over all prefixes (169 runs, April~7--25) and a longitudinal campaign over the same set approximately twice daily through the 26~May restoration. To characterize the anomalous port-5353 population (Section~\ref{sec:census}), we additionally probe all 256 addresses in 20 affected \texttt{/24} blocks on UDP/5353 using a standard mDNS/DNS-SD enumeration query, with TCP/80 and TCP/22 as controls, and repeat the probes with a random nonexistent query name to test response selectivity. Results are reported in Section~\ref{ssec:injection}.

\paragraph{Ethics.}
All probes are rate-limited TCP SYN attempts that impose negligible load and originate from identified vantage points with opt-out webpages. We report only aggregate AS- and prefix-level results, publish no individual addresses, particularly those of identifiable consumer devices on the shutdown floor, and retain no user-identifying data.
\section{Censys Host Visibility}
\label{sec:censys}
This section characterizes how the 2026 shutdowns appear in Internet-wide scan data and shows that this view can be misleading in specific, quantifiable ways. Censys host counts capture both shutdown onsets clearly but diverge from actual connectivity in three respects. First, visibility does not return to its pre-shutdown baseline after either restoration; instead, it increases to roughly $3.6\times$ the baseline and remains elevated. Second, mid-shutdown visibility climbs roughly sevenfold during Event~2, weeks before restoration, without any corresponding change in reachability. Third, restoration appears as a three-week ramp, although per-AS decomposition shows that its onset was near-synchronous and only the refill was gradual. We first establish the overall trend and baseline, then examine each divergence in turn.

\begin{figure*}[t!]
\centering
\begin{tikzpicture}
\begin{axis}[
  iranstyle,
  width=\textwidth, height=6.2cm,
  ymode=log,
  ymin=7000, ymax=6000000,
  ytick={10000,30000,100000,300000,1000000,3000000},
  yticklabels={10K,30K,100K,300K,1M,3M},
  ylabel={Censys-visible Iranian hosts (log scale)},
  xmin=2026-01-01, xmax=2026-07-01,
  xtick={2026-01-01,2026-01-15,2026-02-01,2026-02-15,2026-03-01,2026-03-15,
         2026-04-01,2026-04-15,2026-05-01,2026-05-15,2026-06-01,2026-06-15,2026-07-01},
  xticklabels={1 Jan,15 Jan,1 Feb,15 Feb,1 Mar,15 Mar,1 Apr,15 Apr,
               1 May,15 May,1 Jun,15 Jun,1 Jul},
  x tick label style={font=\scriptsize,rotate=35,anchor=east},
  y tick label style={font=\scriptsize},
  grid=major, grid style={gray!20},
]

\fill[shutdownred!10]   (axis cs:2026-01-08,7000) rectangle (axis cs:2026-01-25,6000000);
\fill[shutdownred!10]   (axis cs:2026-03-01,7000) rectangle (axis cs:2026-05-26,6000000);

\fill[gray!18] (axis cs:2026-01-01,910000) rectangle (axis cs:2026-07-01,1170000);
\draw[gray!60,dashed,line width=0.6pt] (axis cs:2026-01-01,1170000) -- (axis cs:2026-07-01,1170000);
\draw[gray!60,dashed,line width=0.6pt] (axis cs:2026-01-01,910000)  -- (axis cs:2026-07-01,910000);
\node[black!70,font=\scriptsize,anchor=west] at (axis cs:2026-04-05,1030000) {2025 baseline};

\addplot[color=lineblue, line width=1.0pt, mark=*, mark size=0.7pt,
         mark options={fill=lineblue}] coordinates {
  (2026-01-01,929657)(2026-01-02,926327)(2026-01-03,907999)(2026-01-04,908581)(2026-01-05,938220)
  (2026-01-06,935290)(2026-01-07,935833)(2026-01-08,1175817)(2026-01-12,46256)(2026-01-13,88054)
  (2026-01-14,41879)(2026-01-15,41204)(2026-01-16,40207)(2026-01-17,62300)(2026-01-18,62080)
  (2026-01-19,43997)(2026-01-20,68617)(2026-01-21,39102)(2026-01-22,35172)(2026-01-23,28609)
  (2026-01-24,18991)(2026-01-25,183907)(2026-01-26,260317)(2026-01-27,739530)(2026-01-28,1365649)
  (2026-01-29,2100155)(2026-01-30,2558559)(2026-01-31,2869443)(2026-02-01,3083597)(2026-02-02,3200092)
  (2026-02-03,3234575)(2026-02-04,3258632)(2026-02-05,3273823)(2026-02-06,3288995)(2026-02-07,3293203)
  (2026-02-08,3318276)(2026-02-09,3325595)(2026-02-10,3330857)(2026-02-11,3329904)(2026-02-12,3329202)
  (2026-02-13,3332434)(2026-02-14,3336868)(2026-02-15,3357377)(2026-02-16,3370098)(2026-02-17,3372908)
  (2026-02-18,3391254)(2026-02-19,3405478)(2026-02-20,3410410)(2026-02-21,3398724)(2026-02-22,3410070)
  (2026-02-23,3424094)(2026-02-24,3451278)(2026-02-25,3470394)(2026-02-26,3481772)(2026-02-27,3477204)
  (2026-02-28,3469568)(2026-03-01,3396330)(2026-03-02,402466)(2026-03-03,110141)(2026-03-04,65973)
  (2026-03-05,63966)(2026-03-06,62980)(2026-03-07,60595)(2026-03-08,58885)(2026-03-09,55361)
  (2026-03-10,73074)(2026-03-11,44344)(2026-03-12,40992)(2026-03-13,36332)(2026-03-14,28664)
  (2026-03-15,23012)(2026-03-16,60977)(2026-03-17,81348)(2026-03-18,9873)(2026-03-19,45380)
  (2026-03-20,53362)(2026-03-21,18846)(2026-03-22,10850)(2026-03-23,11145)(2026-03-24,34932)
  (2026-03-25,11533)(2026-03-26,11762)(2026-03-27,11772)(2026-03-28,11497)(2026-03-29,11343)
  (2026-03-30,10990)(2026-03-31,32533)(2026-04-01,10791)(2026-04-02,10738)(2026-04-03,10721)
  (2026-04-04,10873)(2026-04-05,11132)(2026-04-06,11538)(2026-04-07,31946)(2026-04-08,21590)
  (2026-04-09,28564)(2026-04-10,36162)(2026-04-11,42695)(2026-04-12,52595)(2026-04-13,54472)
  (2026-04-14,62030)(2026-04-15,61276)(2026-04-16,62385)(2026-04-17,63173)(2026-04-18,63558)
  (2026-04-19,64038)(2026-04-20,62855)(2026-04-21,62744)(2026-04-22,63168)(2026-04-23,63909)
  (2026-04-24,64127)(2026-04-25,64185)(2026-04-26,64380)(2026-04-27,64864)(2026-04-28,64810)
  (2026-04-29,64595)(2026-04-30,65213)(2026-05-01,65712)(2026-05-02,65587)(2026-05-03,65822)
  (2026-05-04,65648)(2026-05-05,66183)(2026-05-06,66790)(2026-05-07,66804)(2026-05-08,67066)
  (2026-05-09,67299)(2026-05-10,67554)(2026-05-11,67902)(2026-05-12,68486)(2026-05-13,68889)
  (2026-05-14,68621)(2026-05-15,69530)(2026-05-16,70660)(2026-05-17,70872)(2026-05-18,71883)
  (2026-05-19,72888)(2026-05-20,72718)(2026-05-21,70196)(2026-05-22,70461)(2026-05-23,70752)
  (2026-05-24,71274)(2026-05-25,71436)(2026-05-26,72652)(2026-05-27,573669)(2026-05-28,982601)
  (2026-05-29,1331427)(2026-05-30,1626932)(2026-05-31,1884928)(2026-06-01,2106704)(2026-06-02,2296560)
  (2026-06-03,2462548)(2026-06-04,2586112)(2026-06-05,2699873)(2026-06-06,2803521)(2026-06-07,2875821)
  (2026-06-08,2945573)(2026-06-09,3015718)(2026-06-10,3072399)(2026-06-11,3143129)(2026-06-12,3225279)
  (2026-06-13,3265232)(2026-06-14,3307384)(2026-06-15,3336874)(2026-06-16,3357679)(2026-06-17,3354085)
  (2026-06-18,3358602)(2026-06-19,3357291)(2026-06-20,3362978)(2026-06-21,3378234)(2026-06-22,3382408)
  (2026-06-23,3386302)(2026-06-24,3384575)(2026-06-25,3351861)(2026-06-26,3345135)(2026-06-27,3360957)
  (2026-06-28,3358451)(2026-06-29,3357823)(2026-06-30,3342245)
};

\draw[shutdownred,dashed,line width=0.8pt] (axis cs:2026-01-08,7000) -- (axis cs:2026-01-08,6000000);
\node[shutdownred,font=\scriptsize\bfseries,rotate=90,anchor=south west]
  at (axis cs:2026-01-08,8000) {Event~1 onset};
\draw[shutdownred,dashed,line width=0.8pt] (axis cs:2026-03-01,7000) -- (axis cs:2026-03-01,6000000);
\node[shutdownred,font=\scriptsize\bfseries,rotate=90,anchor=south west]
  at (axis cs:2026-03-01,8000) {Event~2 onset};

\draw[shutdowngreen,dashed,line width=0.8pt] (axis cs:2026-01-25,7000) -- (axis cs:2026-01-25,6000000);
\node[shutdowngreen,font=\scriptsize\bfseries,rotate=90,anchor=south west]
  at (axis cs:2026-01-25,80000) {Restoration 1};
\draw[shutdowngreen,dashed,line width=0.8pt] (axis cs:2026-05-26,7000) -- (axis cs:2026-05-26,6000000);
\node[shutdowngreen,font=\scriptsize\bfseries,rotate=90,anchor=south west]
  at (axis cs:2026-05-26,80000) {Restoration 2};

\node[anomalypurple,font=\scriptsize,anchor=south] at (axis cs:2026-02-14,3500000)
  {plateau (3.48M)};
\node[anomalypurple,font=\scriptsize,anchor=south] at (axis cs:2026-06-14,3500000)
  {plateau (3.34M)};

\node[black!65,font=\scriptsize,anchor=north east] at (axis cs:2026-05-20,130000)
  {mid-shutdown climb ($7\times$)};
\node[black!65,font=\scriptsize,anchor=north] at (axis cs:2026-03-28,11500)
  {floor ${\approx}10$--11K};

\end{axis}
\end{tikzpicture}
\caption{Censys-visible Iranian IPv4 hosts, 1~January--30~June 2026 (178 daily observations; 9--11~January are absent from the dataset). \emph{Log vertical scale}, spanning the 9{,}873 shutdown floor to the ${\approx}3.4$M plateaus. Red bands and dashed lines mark the two shutdowns (onsets 8~January and 1~March); green dashed lines mark the two restorations.}

\label{fig:iran-host-timeseries}
\end{figure*}
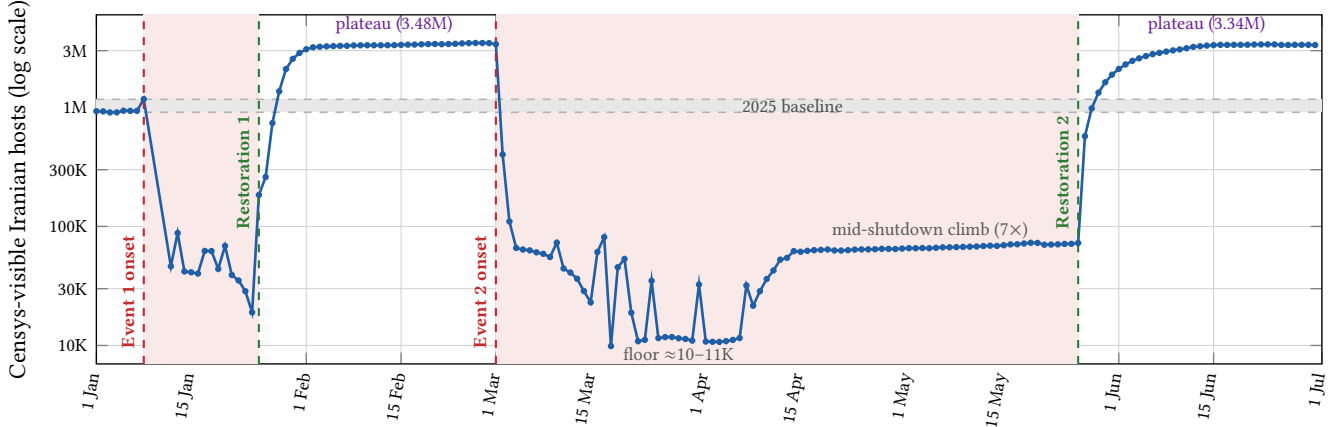

Figure~\ref{fig:iran-host-timeseries} shows the daily number of Censys-visible IPv4 hosts in Iran from January~1 to June~30, 2026 (178 observations; January 9--11 are absent from the dataset). The counts include all hosts present in each snapshot, regardless of pending-eviction state. Because Censys retains unresponsive hosts for up to 72~hours before removal~\cite{durumeric2025censys}, these raw counts may overstate reachability during rapid enforcement events. We therefore perform a host-level analysis on key dates, distinguishing \emph{active} hosts. From May~15, 2026 onward, the Censys dataset reports no pending hosts for Iran. This behavior is not specific to Iran, as our Turkey control, which experienced no disruption, likewise reports 8--9\% pending hosts through March~18 and exactly 0\% from late May. We therefore treat the pending decomposition as \emph{unavailable} after May~14 rather than interpreting the reported value as a true zero. Accordingly, we restrict all pending-based analyses to snapshots on or before May 14. Raw host counts are unaffected.

\textbf{Baseline.}
Twelve monthly snapshots from 2025 establish a stable baseline of ${\sim}910$K--1.17M hosts. The pre-shutdown snapshot from January~2026 contains ${\sim}935$K hosts, which falls within this range. At this point, 14--17\% of hosts are in the eviction queue, reflecting normal pipeline churn. We therefore use the January~2026 host count (${\sim}935$K) as the reference denominator. Table~\ref{tab:onset-summary} summarizes the host-level decomposition at key dates for both shutdown events.

\textbf{Overall trend.}
The six-month series shows a recurring two-shutdown structure. Each event has four phases, namely a sharp collapse, a sustained floor, a restoration, and an elevated post-restoration plateau. Event~1 begins on January~8, when the visible host count falls from ${\sim}935$K to a floor that deepens to ${\sim}19$K by January~24 ($0.02\times$ baseline). Restoration occurs on January~25, after which visibility does not return to the pre-shutdown baseline. Instead, it rises over six days to a plateau of approximately $3.3$--$3.5$M hosts that holds for a month, and peaks at 3.34M on February~26 ($3.6\times$ baseline). Event~2 begins on March~1, when 99.1\% of the 3.4M visible hosts enter the eviction queue in a single day, leaving only 30K actively responding hosts. The reported host count then declines to ${\sim}66$K within three days and reaches a sustained floor of ${\sim}10$--11K through late March and early April. Restoration is announced on May~26 and is visible the same day in the forwarding plane (Section~\ref{sec:census}). Censys visibility then recovers over approximately three weeks to a plateau of ${\sim}3.34$M hosts ($3.6\times$ baseline), which persists through June~30.

Two features recur across both events and organize the remainder of this section. First, each restoration is followed by an overshoot to ${\sim}3.6\times$ the baseline (Sections~\ref{ssec:inflation} and \ref{ssec:inflation-dist}). Second, each shutdown floor retains a small, stable residual host population (Section~\ref{ssec:floor}). A third feature is unique to Event~2, with the visible host count rising from the ${\sim}11$K April~6 floor to ${\sim}73$K by May~19, weeks before any restoration (Section~\ref{ssec:climb}).

\begin{table}[t]
\centering
\caption{Censys total vs.\ active Iranian host counts at key dates. Ratios are active hosts relative to the ${\approx}935K$ January baseline. $^\dagger$From May~15 the pending field is unpopulated dataset-wide (see text), so ``active'' equals the raw count and ratios are upper bounds.}
\small
\setlength{\tabcolsep}{4pt}
\label{tab:onset-summary}
\renewcommand{\arraystretch}{0.95}
\begin{tabular}{@{}llrrrr@{}}
\toprule
Date & Phase & Total & Active & Pend.\ \% & Act./base \\
\midrule
Jan~7  & baseline        & 936K  & 798K  & 14.8 & $0.85\times$ \\
Jan~8  & onset           & 1.18M & 1.02M & 13.2 & $1.09\times$ \\
Jan~12 & floor           & 46K   & 44K   &  4.1 & $0.05\times$ \\
Jan~24 & deep floor      & 19K   & 16K   & 16.0 & $0.02\times$ \\
\addlinespace
Jan~25 & restoration 1   & 184K  & 178K  &  3.4 & $0.19\times$ \\
Jan~27 & ramp            & 740K  & 695K  &  6.0 & $0.74\times$ \\
Feb~26 & peak            & 3.48M & 3.34M &  4.2 & $3.57\times$ \\
\midrule
Feb~28 & pre-onset       & 3.47M & 3.16M &  9.0 & $3.38\times$ \\
Mar~1  & true onset      & 3.40M & 30K   & 99.1 & $0.03\times$ \\
Mar~2  & Censys onset    & 402K & 56K & 86.1 & $0.06\times$ \\
Mar~18 & deepest day     & 10K   & 4K    & 62.5 & $0.004\times$ \\
Apr~1  & sustained floor & 11K   & 10K   & 11.8 & $0.01\times$ \\
\addlinespace
Apr~7  & climb begins    & 32K   & 30K   &  5.5 & $0.03\times$ \\
May~26 & restoration 2   & 73K   & 73K   & n/a\rlap{$^\dagger$} & $0.08\times$ \\
May~27 & refill day 1    & 574K  & 574K  & n/a  & $0.61\times$ \\
May~31 & refill          & 1.88M & 1.88M & n/a  & $2.01\times$ \\
Jun~30 & plateau         & 3.34M & 3.34M & n/a  & $3.57\times$ \\
\bottomrule
\end{tabular}
\end{table}

\subsection{Shutdown Onset}
\label{ssec:onset} 
Both events begin with a collapse that is simultaneous across operator categories, but the two onsets are recorded very differently by the scanning pipeline, and only one of them is dated correctly by the raw host count.

The collapse is broadly uniform across independently operated networks. In Event~1, visible hosts fall by comparable proportions across AS categories between January~7 and January~16. State telecom declines by 98.7\% (101K to 1.3K), commercial ISPs by 95.5\% (727K to 33K), and mobile infrastructure by 94.4\% (3.6K to 0.2K), a spread of only 4.3 pp across three independently operated categories. The collapse is again broadly uniform in Event~2, with one exception. 
Iran Telecommunication Company (TCI-AS58224) falls by 99.8\% (1.34M to 2{,}368) and every remaining category by more than 95\%, except mobile infrastructure (92.6\%), whose smaller decline reflects MCCI's comparatively modest drop (77.9\%, 10{,}130 to 2{,}237). The near-synchronous collapse of networks under separate operational control is consistent with centralized enforcement at TIC's international gateway rather than with independent action by individual operators.

Event~2’s onset is visible in the eviction queue a day before it appears in the raw host count. Because Censys retains unresponsive hosts for up to 72~hours, the raw count lags rapid enforcement, while the pending decomposition identifies the onset more precisely. On February~28, 231K hosts become newly pending, $4.2\times$ the preceding four-day average of 55K/day, consistent with partial enforcement one day before the full shutdown. On March~1, 99.1\% of the 3.40M visible hosts are in the eviction queue, while only 30K remain actively responding, establishing March~1 as the onset of the observed disruption and yielding a visible host count of just $0.03\times$ the baseline. The raw count does not reflect this collapse until March~2, when it stands at 402K, 86\% of which is carry-over from March~1, and does not approach the floor until March 3--4. A study relying only on the raw host-count series would place the onset a day or two late and understate the abruptness of the collapse.

Event 1’s onset is bounded by a gap in the data. The January~8 snapshot shows 1.02M active hosts with normal churn, with 13.2\% pending compared with 14–17\% at baseline, indicating that enforcement had not yet taken effect at scan time. January 9–11 are absent from the dataset, and by January~12 the count has fallen to 46K, with only 4.1\% pending, indicating that the collapse had already occurred and the eviction queue had largely drained. The onset therefore falls within the January 8–12 window and, based on reports from that period~\cite{ioda-comparative}, is placed on the evening of January~8. The low pending fraction on January~12 further indicates that the observed collapse is not a retention artifact but reflects a disruption that had already fully propagated.

\subsection{Post-Restoration Inflation}
\label{ssec:inflation}
Neither restoration returns Iranian host visibility to its pre-shutdown level; both are followed by a plateau at ${\sim}3.6\times$ the baseline that persists throughout our observation. After Restoration~1 (25~January), visibility climbs over six days to ${\sim}3.3$M and remains elevated for a month, peaking at ${\sim}3.34$M on February~26 ($3.6\times$ the ${\sim}935$K baseline). After Restoration~2 (26~May), visibility refills over three weeks to ${\sim}3.34$M and remains at this level through 30~June. The two plateaus are quantitatively similar despite following restorations four months apart and shutdowns of very different durations (17 vs. 86 days). The first persists for a month until Event~2 interrupts it, while the second lasts at least six weeks, indicating a durable change rather than a transient restoration artifact.

The inflation is specific to Iran. On February~26, when Iran peaked at 3.34M hosts, our Turkey control stood at 1.30M, within its normal range; on June~15, Iran was at 3.34M and Turkey at 1.53M. Turkey varies only between ${\sim}1.30$M and ${\sim}1.53$M over the period, while the same dates in 2025 show 1.63M--1.82M with no comparable inflation. Thus, a comparable undisrupted country remains stable while Iranian visibility triples, ruling out a global change in scanning intensity.

The inflation is not pending-eviction residue of Censys. At the February~26 peak, only 4.2\% of hosts are pending removal, compared with 14–17\% at baseline, and the active-host filter excludes only 1.6–4.7\% across inter-event dates. The inflated addresses respond to scanning, but active probing in Section~\ref{ssec:injection} shows these responses are synthesized by a middlebox at Iran's international gateway rather than emitted by genuine hosts. The ${\sim}3.6\times$ overshoot therefore does not reflect a change in Iran's reachable host population.

This has a direct measurement consequence: Censys host counts are not a reliable measure of Iranian connectivity or shutdown severity. Reading this series as a connectivity signal would imply that Iran became $3.6\times$ better connected immediately after each shutdown, contradicting both our forwarding-plane measurements and the injection finding in Section~\ref{sec:census}. Section~\ref{ssec:inflation-dist} decomposes the inflation across address space and operators.

\subsection{Anatomy of the Inflation}
\label{ssec:inflation-dist}
Section~\ref{ssec:inflation} established that both restorations are followed by a ${\sim}3.6\times$ overshoot that is specific to Iran and sustained. Here we decompose it along two axes, address space and operator, to locate where the additional hosts appear.

The inflation reflects depth rather than breadth. Table~\ref{tab:geometry} decomposes four representative snapshots by the number of visible hosts, distinct \texttt{/24} blocks, and distinct origin ASes. Between the January baseline and the June plateau, the host count increases by $3.6\times$, but the address footprint changes little. The distinct \texttt{/24} blocks increase by only $1.16\times$ (17,563 to 20,414), while distinct origin ASes decrease slightly (554 to 543). The increase is therefore almost entirely absorbed as density: hosts per /24 rise from 53.4 to 163.5. Thus, Censys is not discovering new Iranian networks after the restoration, and it is observing roughly three times as many hosts within the same ${\sim}20$K /24 blocks and ${\sim}550$ ASes.

The inflation is also concentrated on a single port. At both plateaus, ${\approx}89\%$ of visible hosts expose port~5353 (mDNS) at a 100\% empty-banner rate (3.09M of 3.48M on February~26; 2.97M of 3.34M on June~15), against 6.7\% at the January baseline, while genuine service ports (HTTP, HTTPS, SSH) carry real banners and stay flat. The overshoot is thus almost entirely one empty-response population, which Section~\ref{ssec:injection} identifies by active probing as response injection~\cite{bock2021weaponizing}.

\begin{table}[t]
\centering
\caption{Address-space geometry of Iranian host visibility at four representative dates. Host counts rise $3.6\times$ from baseline to plateau while the \texttt{/24} and AS footprint is essentially unchanged; the increase is absorbed entirely as density.}
\label{tab:geometry}
\small
\setlength{\tabcolsep}{3.2pt}
\renewcommand{\arraystretch}{1.1}
\begin{tabular}{@{}llrrrrr@{}}
\toprule
Date & Phase & Hosts & \texttt{/24}s & ASes
  & \makecell{Hosts\\per \texttt{/24}} & \makecell{Hosts\\per AS} \\
\midrule
Jan~5  & baseline        &  938K & 17{,}563 & 554 &  53.4 & 1{,}694 \\
Feb~26 & inflation~1     & 3.48M & 20{,}692 & 580 & 168.3 & 6{,}003 \\
Apr~1  & shutdown floor  &   11K &  1{,}574 & 217 &  6.9 &   50 \\
Jun~15 & inflation~2     & 3.34M & 20{,}414 & 543 & 163.5 & 6{,}145 \\
\bottomrule
\end{tabular}
\end{table}

The shutdown, by contrast, reduces both breadth and depth. At the April floor, visible \texttt{/24} blocks fall to 2,182 (12\% of baseline) and origin ASes to 261 (47\%). Enforcement removes whole blocks from view; the subsequent inflation does not restore them but thickens the blocks that remain. The two phenomena are geometrically distinct: the inflation is not a ``recovery beyond baseline.''

The two inflations exhibit the same address-space regime. February~26 and June~15 are nearly identical on every measure (3.48M vs.\ 3.34M hosts, 20,692 vs.\ 20,414 \texttt{/24}s, 580 vs.\ 543 ASes), despite occurring four months apart following 17- and 86-day shutdowns. The reproducibility of this pattern suggests that the overshoot is governed by a consistent mechanism with a well-defined ceiling rather than by restoration-specific effects.

The depth increase is largest in academic networks by \emph{multiplier}, not by mass. Table~\ref{tab:multipliers} decomposes the second inflation by AS. Backbone, mobile, and access operators grow at or below the country-wide $3.6\times$ (TCI $4.3\times$, Respina $4.5\times$, Afranet $1.2\times$), while academic networks grow by one to three orders of magnitude (IRANET-IPM $129\times$, University of Tehran $391\times$, TUMS $401\times$). But these ratios, computed from baselines of tens of hosts, measure relative amplification, not absolute mass: TCI alone adds ${\approx}1.08$M of the ${\approx}2.4$M excess, the academic ASes together under 7\%. Both observations reflect the same phenomenon. The port-5353 signature is ${\approx}89\%$ of \emph{all} plateau hosts (Section~\ref{ssec:inflation-dist}), including TCI, while academic networks exhibit it most clearly because they contain the most dark address space for the injector (Section~\ref{ssec:injection}) to populate.

\begin{table}[t]
\centering
\caption{Per-AS host counts at the pre-shutdown baseline (5~Jan) and the
post-restoration plateau (15~Jun), ordered by plateau size. Operators already
well-observed at baseline grow at or below the country-wide $3.6\times$;
academic and research networks, nearly invisible at baseline, grow by one to
three orders of magnitude. $^\dagger$Ratios against baselines of tens of hosts
are indicative only.}
\label{tab:multipliers}
\small
\setlength{\tabcolsep}{7pt}
\renewcommand{\arraystretch}{1.1}
\begin{tabular}{@{}rlrrr@{}}
\toprule
ASN & Operator & Category & \makecell[r]{Jan~5} & \makecell[r]{Jun~15\\($\times$)} \\
\midrule
\multicolumn{5}{@{}l}{\textit{Already well-observed at baseline}}\\
58224  & TCI            & state    & 323{,}731 & 1.40M ($4.3$) \\
42337  & Respina        & comm.    &  50{,}573 & 228K ($4.5$) \\
206065 & FDI            & comm.    &  92{,}320 & 155K ($1.7$) \\
50810  & Mobinnet       & mobile   &  49{,}368 &  60K ($1.2$) \\
25184  & Afranet        & comm.    &  41{,}076 &  48K ($1.2$) \\
43754  & Asiatech       & comm.    &  28{,}819 &  34K ($1.2$) \\
31549  & Rasana         & comm.    &  28{,}308 & 239K ($8.4$) \\
\midrule
\multicolumn{5}{@{}l}{\textit{Sparsely observed at baseline}}\\
24631  & FanapTelecom   & comm.    &   6{,}099 & 101K ($16.5$) \\
16322  & ParsOnline     & comm.    &   3{,}969 &  59K ($15.0$) \\
12880  & DCI            & state    &   1{,}095 &  20K ($18.4$) \\
197207 & MCCI           & mobile   &   2{,}903 &  16K ($5.6$) \\
\midrule
\multicolumn{5}{@{}l}{\textit{Academic and research}}\\
6736   & IRANET-IPM     & academic &       577 &  75K ($129$)$^\dagger$ \\
12660  & Sharif Univ.   & academic &       748 &  19K ($25$) \\
29068  & Univ.\ Tehran  & academic &        55 &  21K ($391$)$^\dagger$ \\
43965  & TUMS           & academic &        51 &  20K ($401$)$^\dagger$ \\
57563  & Isfahan Med.   & academic &        37 &   9K ($248$)$^\dagger$ \\
59506  & Shiraz Med.    & academic &         9 &  13K ($1434$)$^\dagger$ \\
\bottomrule
\end{tabular}
\end{table}

The same networks drive both inflations. Applying the decomposition to the first event, academic ASes grow $16\times$ between January~7 and March~1 ( ASes grow $16\times$ between January~7 and March~1 (${\approx}13$K to ${\approx}214$K), compared with a $3.6\times$ overall rise, led by the same operators at nearly identical factors (IRANET-IPM $127\times$ and  University of Tehran $388\times$, versus $129\times$ and $391\times$ in the second inflation). The same networks also drive the transient March~16--17 increase (Section~\ref{ssec:floor}), which is likewise injected rather than a genuine recovery: the injector targets the same academic address space in every episode.

The geometry rules out benign explanations. The footprint is unchanged, so the inflation is not new address space coming online, while the Turkey control remains flat over the same dates (Section~\ref{ssec:inflation}), ruling out a global scanning change. The remaining explanation is a change in how Censys observes a fixed set of Iranian networks, which Section~\ref{ssec:injection} directly attributes to UDP/5353 response injection by middleboxes rather than any change in the host population.

\subsection{The Shutdown Floor}
\label{ssec:floor}
Both events reach a visibility floor, but the Event~2 floor is deeper, longer-lasting, and more stable. After the initial collapse associated with the eviction queue (Section~\ref{ssec:onset}), visibility settles at ${\approx}10$--11K hosts from 22 March through 6 April, representing roughly 1\% of the ${\sim}935$K baseline and the lowest sustained visibility level observed in our six-month measurement window. In contrast, Event~1’s floor persists for less than two weeks and fluctuates between 19K and 88K hosts, never reaching a comparable reduction in visibility.

The floor is not entirely stable. Transient spikes interrupt it, the largest on March~16--17 (61K and 81K hosts), with smaller increases on March~19--20, 24, and 31. Each lasts one to two days before visibility returns to the floor. The March~16--17 rise is a $+54$K increase over March~15, 79.5\% of it academic (+43K). These are not universities becoming reachable, however, but the response injection of Section~\ref{ssec:injection} briefly expanding into academic address space: in the four networks driving the spike, empty-banner port-5353 responders rise from ${\sim}4{,}500$ to ${\sim}35{,}700$ over two days, while their genuine service ports stay in the low hundreds (HTTP~194, HTTPS~201, SSH~15, with real banners). A small real increase (real-bannered hosts rising from tens to ${\sim}200$) may accompany it but cannot account for the ${\sim}31$K-host increase. Similarly, the spike coincides with no change in BGP prefix visibility (Section~\ref{sec:bgp-routing}): the entire movement is injected visibility, invisible to both the routing and forwarding planes.

The lowest-visibility day is dominated by pending removals. On 18~March, the minimum count of the entire period is 9,873 hosts, but 6,170 (62.5\%) are pending removal, representing the highest pending fraction outside the 1~March enforcement transition. Only ${\approx}3{,}700$ hosts remain actively responding. The pending population is dominated by hosting and telecom networks rather than academic networks, consistent with residual effects from the 16--17 March transient recovery being removed through the retention process. Thus, 18~March represents the trough between two partial recoveries, with the active-host floor reaching roughly one-third of the headline count.

The sustained floor is stable and not explained by measurement artifacts. From 25~March to 6~April, excluding the 31~March bump, counts remain between 10,721 and 11,772 hosts, with pending fractions of 7–18\%, consistent with the 14--17\% observed at the pre-shutdown baseline, and therefore consistent with ordinary pipeline churn rather than a draining eviction queue. This residual population is consistent with active probing: our forwarding-plane census over the same window finds 96.9\% of Iranian prefixes silently discard probes, with a small reachable remainder (Section~\ref{sec:census}). Because that remainder is a lower bound (Section~\ref{sec:method}), the two methods are mutually consistent rather than independently confirming, but together they indicate a stable residual of Iranian infrastructure remained reachable through the deepest phase of the shutdown. Section~\ref{sec:as-analysis} identifies the networks comprising this residual and shows that a substantial fraction were not traversing TIC’s enforcement point.

\subsection{The Mid-Shutdown Climb}
\label{ssec:climb}
During the Event~2 shutdown, the visible host count rises from ${\sim}11$K on April~6 to ${\sim}73$K by May~19, a sevenfold increase that superficially suggests a gradual partial recovery. Instead, the increase comes from the same injected population that dominates the post-restoration plateau, whose visibility expands mid-shutdown.

The climb exhibits the same geometry as the post-restoration inflations (Section\ref{ssec:inflation-dist}). It reflects depth rather than breadth. The address footprint barely moves (distinct \texttt{/24}s fall from 4,192 to ${\sim}2{,}750$ by late April, then recover to 4,491; origin ASes track it, 400 to ${\sim}295$ to 376), while hosts per \texttt{/24} roughly triple. And, as in every other inflation episode, the growth concentrates in academic networks (University of Tehran $176\times$, Shiraz $125\times$, IRANET-IPM $42\times$, Allameh $46\times$) while backbone and infrastructure operators stay flat or decline (MCCI $1.1\times$, ArvanCloud $1.0\times$, Asiatech $0.7\times$). The pending fraction also falls to 1--3\% as the climb proceeds, so the rise is not retention-queue churn.

The climb reflects response injection rather than recovery. Two mechanisms could produce a rise that our per-prefix breadth probe cannot distinguish: hosts genuinely becoming reachable through granular whitelisting, which is reported for this shutdown~\cite{madory2026iranian_shutdown}, or expansion of the response injection identified in Section~\ref{ssec:injection}. We discriminate between them by service composition. At the climb peak, the academic networks driving the increase contain ${\approx}26{,}500$ empty-banner responders on UDP/5353 (100\% empty), compared with only tens of hosts on genuine service ports (HTTP~30, HTTPS~35, SMTP~34, all with valid banners). Whitelisting would add reachable hosts with real service banners; instead, the climb consists almost entirely of injected UDP/5353 responses of the same kind, along with responses on the same secondary ports (161/SNMP, likewise 100\% empty), that dominate the plateau. The mid-shutdown climb therefore reflects expansion of the injector's footprint within these networks, not a policy-driven reachability change, consistent with the flat $\sim$2\% forwarding-plane reachability observed throughout the window (Section~\ref{sec:census}). We cannot exclude a small genuinely reachable population among the tens of hosts with real banners, but it cannot account for the sevenfold rise. A passive-scan-only study would interpret this period as gradual mid-shutdown recovery, but the forwarding-plane and service-composition evidence shows otherwise.

\section{Survivor Networks and Routing Paths}
\label{sec:as-analysis}
Passive scanning identifies which networks remain ``visible'' during a shutdown, but visibility conflates two mechanisms that a shutdown study must distinguish. Some networks genuinely survive enforcement while routed through Iran, whereas other address space only geolocates to Iran but is routed through foreign upstreams and was never subject to TIC's enforcement. Because Iranian BGP announcements remain globally present throughout the 2026 shutdowns (Section~\ref{sec:bgp-routing}), control-plane presence cannot distinguish these cases. The same decoupling that hides the shutdown from BGP monitors also obscures survivor attribution. We resolve this ambiguity by combining the composition of the shutdown floor with per-AS routing-path classification. We assign AS categories using the ipverse metadata dataset~\cite{ipverse}, supplemented by keyword matching and manual overrides for ASes with null or ambiguous categories.

The persistent survivors of the deep shutdown floor are infrastructure, not
access networks. Across the shutdown floor (22~March--6~April, ${\sim}$10--11K visible hosts), the hosts present on at least 12 of the 14 floor days are dominated not by end-user access networks but by infrastructure: the mobile core (MCCI, ${\sim}$2{,}500 persistent hosts, the largest survivor), hosting and datacenter networks, the state backbone (DCI, TCI), and TIC's own gateway (AS49666). At the deepest day (18~March, ${\sim}$3{,}700 genuinely active hosts once ${\sim}$6{,}170 pending-eviction hosts are excluded), this concentration is even starker: three CDN/cloud/hosting networks account for over 60\% of the live floor, while the mobile core and state telecom are themselves largely mid-collapse (MCCI: 10 active; TCI: 183 of 479). Academic networks are effectively absent from the floor. The University of Tehran and Sharif each have zero hosts for at least 12 days, confirming that their apparent visibility on 16--17~March (Section~\ref{ssec:floor}) was injected rather than genuine. These networks host only tens of reachable end systems even as thousands of injected port-5353 responders appear in their address space (Section~\ref{ssec:injection}). As the routing analysis below shows, several of these apparent survivors (\eg ArvanCloud) were never behind TIC at all.

Routing paths separate genuine survivors from address space that only geolocates
to Iran. To distinguish the two survival mechanisms, we parsed RIPE RIS bview snapshots for five dates (baseline 2026-01-05; floor 2026-03-18, 03-25, 04-01; post-restoration 2026-06-15) and, for each candidate AS, classified every observed AS-path as ``via Iran'' if it traversed the Iranian transit set (AS49666, AS48159, AS6736), reporting the per-peer fraction of paths crossing Iran. Five known TIC-downstream networks served as controls and validate the method: all show a via-Iran fraction of ${\approx}1.00$ on every date. Among the flagged survivors, the result is decisive (Table~\ref{tab:survivors}). Four networks, namely M247 (global transit), Batterflyai (Russia-registered), SGHL1, and FrostyHosting, did not route via Iran on any date, including the pre-shutdown baseline, and an independent ASN-to-country check confirms no Iranian AS on any of their paths. These are foreign-hosted networks that only geolocate to Iran. Thus, their ``survival'' is not evidence of exemption. ArvanCloud is reached ${\sim}98\%$ via foreign anycast points of presence (via-Iran ${\approx}0.02$, stable across all dates), the clearest case of anycast-based resilience.

\begin{table}[t]
\centering
\small
\caption{Shutdown-floor survivors (22~Mar--6~Apr), grouped by whether their traffic traverses Iran's enforcement point. \emph{Persistent hosts}: hosts present on ${\ge}12$ of 14 floor days. \emph{Via-Iran fraction}: share of observed RIS paths crossing the Iranian transit set (AS49666/48159/6736). \emph{Verdict}: \checkmark\ TIC-downstream, $\sim$ mixed, \ding{55}\ foreign-routed.}
\label{tab:survivors}
\setlength{\tabcolsep}{8pt}
\renewcommand{\arraystretch}{1.05}
\begin{tabular*}{\columnwidth}{@{\extracolsep{\fill}}lrcc@{}}
\toprule
Operator & \makecell[r]{Persistent\\hosts} & \makecell{Via-Iran\\fraction} & Verdict \\
\midrule
\multicolumn{4}{@{}l}{\textit{TIC-downstream}}\\
\hspace{0.5em}MCCI          & 2{,}484 & 1.00 & \checkmark \\
\hspace{0.5em}HostIran      & 1{,}228 & 1.00 & \checkmark \\
\hspace{0.5em}TCI           & 418     & 1.00 & \checkmark \\
\hspace{0.5em}DCI           & 319     & 1.00 & \checkmark \\
\hspace{0.5em}TIC-GW        & 186     & 1.00 & \checkmark \\
\hspace{0.5em}Irancell      & 69      & 1.00 & \checkmark \\
\addlinespace[2pt]
\multicolumn{4}{@{}l}{\textit{Mixed / partial bypass}}\\
\hspace{0.5em}SabaIdea      & 779 & 0.77       & $\sim$ \\
\hspace{0.5em}Vunify        & 18  & 0.25--0.66 & $\sim$ \\
\addlinespace[2pt]
\multicolumn{4}{@{}l}{\textit{Foreign-routed}}\\
\hspace{0.5em}ArvanCloud    & 1{,}024 & 0.02 & \ding{55} \\
\hspace{0.5em}SGHL1         & 512     & 0.00 & \ding{55} \\
\hspace{0.5em}M247          & --      & 0.00 & \ding{55} \\
\hspace{0.5em}Batterflyai   & 74      & 0.00 & \ding{55} \\
\hspace{0.5em}FrostyHosting & 40      & 0.00 & \ding{55} \\
\bottomrule
\end{tabular*}
\end{table}

Two operators fall between these extremes, showing partial, prefix-specific
bypass rather than a clean verdict. SabaIdea (AS44932), an Iranian content operator (Aparat/Filimo), routes 76.5\% of its observed paths via TIC while the remaining 23.5\% exit through a single foreign upstream, CDN77/Datacamp (AS60068), a UK-based anycast CDN outside TIC's gateway. This foreign slice is not noise. It appears at all 43 RIS peers, converges on that one upstream, and is confined to 5 of SabaIdea's 15 prefixes; the remaining prefixes route domestically through TIC and were subject to the shutdown. The five CDN-routed prefixes ride globally anycast address space that TIC cannot null-route without disrupting the CDN for all its customers, so a defined subset of SabaIdea's footprint stayed reachable while the majority did not, yielding a footprint whose observed survival is proportional to the share of address space routed off TIC. Vunify (AS210814) shifts across the event (via-Iran 0.66${\to}$0.25${\to}$0.51), consistent with mixed reliance on an Iranian upstream (Asiatech) alongside foreign transit; given only five snapshots from a single collector, we do not overinterpret this.

Taken together, these results show that scan visibility overstates in-country
survival. The genuine TIC-downstream survivors are operational infrastructure, including mobile-core signalling, the state backbone, and TIC's own gateway, that remained reachable for routing and internal functions. However, a substantial share of apparent ``Iranian survivors'' in passive scan data consists of foreign-routed address space that was never behind the enforcement point. Censys observes a host whenever it is reachable from its scanners, including when its prefix is routed through foreign upstreams. Scan visibility therefore overstates in-country survival, and distinguishing genuine survival from geolocation artifacts requires routing-plane analysis. SabaIdea further shows that survival is not binary but can be partial and prefix-specific, depending on how much of a network's address space is routed outside TIC.

\section{BGP Routing During Shutdowns}
\label{sec:bgp-routing}
\begin{figure}[t]
\centering
\begin{tikzpicture}
\begin{axis}[
  width=\columnwidth, height=4cm,
  ymin=50, ymax=92,
  ytick={50,55,60,65,70,75,80,85,90},
  ylabel={BGP coverage (\%)},
  xtick={0,1,2,3,4},
  xticklabels={Baseline,Onset,Deep,Restoration,Post},
  x tick label style={font=\footnotesize},
  y tick label style={font=\footnotesize},
  ymajorgrids, grid style={gray!20},
  legend style={font=\footnotesize, at={(0.5,1.03)}, anchor=south,
                legend columns=4, draw=none, fill=none},
]

\fill[red!5] (axis cs:1.65,50) rectangle (axis cs:2.35,92);

\addplot[color=red!60!black, mark=*, line width=1.6pt] coordinates {
  (0,85.1)(1,85.3)(2,54.7)(3,79.9)(4, 85.2)
};
\addlegendentry{2019 (withdrawal)}

\addplot[
  color=blue!70!black,
  mark=square*,
  dashed,
  line width=1.2pt,
  mark size=2.0pt
 ] coordinates {
  (0,88.5)(1,86.9)(2,86.8)(3,88.2)(4,88.3)
};
\addlegendentry{2022}

\addplot[
  color=green!50!black,
  mark=triangle*,
  dashdotted,
  line width=1.2pt,
  mark size=2.2pt
 ] coordinates {
  (0,87.6)(1,86.5)(2,82.0)(3,83.2)(4,86.2)
};
\addlegendentry{2026 E1}

\addplot[
  color=violet!75!black,
  mark=diamond*,
  densely dotted,
  line width=1.3pt,
  mark size=2.2pt
 ]  coordinates {
  (0,80.9)(1,86.0)(2,85.0)(3,86.1)(4,87.3)
};
\addlegendentry{2026 E2}

\node[red!75!black, font=\scriptsize, anchor=west] at (axis cs:2.30,60)
  {2019 withdrawal $\rightarrow$54.7\%};
\draw[red!75!black,->,line width=0.5pt] (axis cs:2.35,59) -- (axis cs:2.05,55.5);

\end{axis}
\end{tikzpicture}
\caption{BGP coverage of RIPE-allocated Iranian IPv4 prefixes across four shutdowns, by phase. Only 2019 shows route withdrawal (to 54.7\% at its deepest); 2022 and 2026 events stay within a few points of baseline through shutdown and restoration.}
\label{fig:bgp-coverage}
\end{figure}
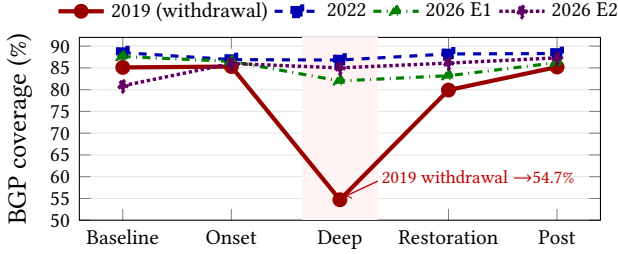

Iran's shutdown mechanism evolves across the four events. In 2019, TIC combines forwarding-plane null routing with partial BGP withdrawal, with coverage dropping from 85.1\% before the shutdown to 54.7\% at the floor, implying ${\approx}440$ prefixes withdrawn from the global routing table. In 2022 and both 2026 events, this hybrid approach is replaced by forwarding-plane null routing alone: TIC discards packets while preserving BGP announcements, making Iran appear reachable to control-plane monitors throughout the shutdowns and their restorations. Figure~\ref{fig:bgp-coverage} summarizes RIPE-allocated Iranian IPv4 prefix coverage across all four events by phase.

Baseline BGP coverage of Iranian address space is stable and high. Of 1,453--1,917 RIPE-allocated Iranian IPv4 prefixes per date, approximately 80–88\% are visible in the global BGP table, either as exact matches or through sub-prefix disaggregation; the remainder are domestically routed or held in reserve. This coverage is stable, at 87.6\% on January~5 and 86.0\% at the first post-onset snapshot of Event~2 (March~3, 2026).

For the 2022 protests and both 2026 events, the globally announced Iranian prefix set remains effectively constant across all phases, including onset, floor, and restoration (Figure~\ref{fig:bgp-coverage}). Coverage stays within a few percentage points of baseline while the forwarding and scan planes change sharply. Three transitions make this explicit. The March 16--17 partial recovery (Section~\ref{sec:as-analysis}), marked by a 54K increase in Censys-visible hosts, coincides with a $-0.4$~pp decrease in coverage (85.2\% to 84.8\%), so the signals move in opposite directions with no corresponding control-plane change. The May 26 restoration, when forwarding-plane reachability steps from approximately 2\% to approximately 10.5\% (Section~\ref{sec:census}), changes coverage only from 85.9\% (May~24) to 86.1\% (May~28). The mid-shutdown Censys climb and post-restoration inflation likewise span only 85.2\% to 87.3\% coverage (April~7 to June~15). Thus, large movements in the forwarding and scan planes occur while the control plane remains essentially flat, indicating that both enforcement and restoration operate below BGP.

Our 2022 snapshots are taken at 00:00 UTC (${\approx}$03:30 Iran time), outside the nightly curfew window during which the 2022 disruption was enforced. Continuous monitoring reports that the 2022 curfews were mobile-scoped and produced a small but visible diurnal dip in aggregate routing announcements~\cite{ioda-comparative}, which discrete daily snapshots cannot resolve. Our 2022 result should therefore be interpreted as fixed-line BGP stability outside the curfew window, consistent with, rather than contrary to, that characterization. The sustained stability of the announced prefix set that we observe in 2026 is qualitatively different.

The 2019 event is the exception, and the one that defines the shift. Coverage remains at 85.1--85.3\% through onset (November 10--17), then falls to 62.3\% by November~20 and 54.7\% by November~21 as ${\approx}440$ prefixes, or 36\% of the covered set, are progressively withdrawn during the deep shutdown, before partially recovering to 79.9\% by November~25. Cross-checks with \texttt{rrc03} and \texttt{rrc04} using the same ASN set deviate by ${\leq}1.7$~pp, indicating that the drop is not specific to \texttt{rrc00}. This withdrawal, absent in every later event, marks a transition from hybrid BGP withdrawal in 2019 to forwarding-plane null-routing by 2022, consistent with the consolidation of enforcement at TIC's international gateway.

\section{Active Forwarding-Plane Probing}
\label{sec:census}
We use active probing for two purposes: to detect the forwarding plane restoration (Section~\ref{ssec:fp-restoration}), and to confirm that the anomalous port-5353 population identified in Sections~\ref{ssec:inflation-dist}, \ref{ssec:floor}, and~\ref{ssec:climb} consists of injected responses rather than genuine hosts (Section~\ref{ssec:injection}).

\subsection{Forwarding-plane Restoration}
\label{ssec:fp-restoration}
We probe Iranian prefixes at the forwarding plane from five vantage points (Section~\ref{sec:method}) in two campaigns: a dense \emph{census} over April~7--25, 2026, deep in Event~2, and a \emph{longitudinal} campaign spanning April through late June and the May~26 restoration. Both probe one address per prefix and therefore measure per-prefix forwarding breadth, specifically whether a prefix’s path discards traffic, rather than how many hosts within a prefix are reachable. Because the probes are TCP-only (ports 80/443/179), a silent-discard verdict cannot distinguish a null route from protocol-selective ACL or firewall drops; we therefore report the dominant verdict as \textsc{silent\_discard} without attributing a specific mechanism.

We probe a single fixed address within each prefix, using the same offset for every prefix and applying the same targets identically across all runs and vantage points. The absolute reachable fraction is not a measure of connectivity: a target may be unallocated or may not run a responding service regardless of enforcement, making the reachable fraction a lower bound dominated by target sparsity. When probed identically over time, however, this fixed set of 4,417 addresses captures changes in forwarding behavior, including a uniform timeout signature that remains stable for weeks before changing abruptly. We therefore use it to characterize the enforcement signature during the shutdown and detect restoration, not to quantify how much of Iran reconnected.

During Event~2, \textsc{silent\allowbreak\_discard} dominates, accounting for 96.5--97.4\% of prefixes across the five vantage points, with a cross-vantage spread of 0.9~pp and a per-vantage run-to-run spread of 1.0--2.0~pp over 18 days of continuous probing. No individual prefix's silence is diagnostic, as silence has many possible causes, from host firewalls to absent hosts and filtered ICMP. The aggregate pattern, however, is consistent with centrally applied enforcement rather than per-host or path-dependent effects. A near-total silence rate that is uniform across five topologically diverse paths, remains stable for 18~days, and changes abruptly and simultaneously at restoration supports this interpretation and is consistent with the BGP stability reported in Section~\ref{sec:bgp-routing}.Academic prefixes deviate from near-total discard, showing an elevated rate of ICMP-unreachable responses (13.3\%) rather than silence. Since BGP coverage remains stable throughout Event~2 (Section~\ref{sec:bgp-routing}), these responses indicate that the prefixes remained announced while reachability was disrupted along the forwarding path, rather than withdrawn from the routing system.

The longitudinal campaign probes the same target set approximately twice daily from all five vantage points. Its aggregate response remains flat for weeks, then changes abruptly within a single ${\sim}18$-hour window on May 26, the announced restoration date. The change is nearly identical across all five vantage points (mean pairwise $r=0.998$) and involves the same set of prefixes changing state everywhere (${\approx}430$ prefixes common to all vantages), consistent with a single, centrally coordinated transition rather than a vantage-specific or gradual one. Crucially, this transition is invisible in BGP: coverage changes by only $0.2$~pp over the same dates (Section~\ref{sec:bgp-routing}). The forwarding plane thus captures both the shutdown and its removal, while the control plane registers neither. After restoration, the discard signature persists for the majority of the target set through the end of our observation window. Consistent with our breadth-only design, we interpret this as continued forwarding-plane discard across much of the address space, not as a measure of connectivity magnitude.

Several limitations bound these forwarding-plane results. These verdicts measure breadth, not depth: probing one address per prefix cannot establish how many hosts within a prefix are reachable, and the reachable fraction is a lower bound because an unresponsive or unallocated address is classified as discard. The TCP-only probes cannot localize the discard point or distinguish null routes from ACL or firewall drops. We therefore use the probes to detect changes in per-prefix path state, most importantly the abrupt, BGP-invisible restoration, rather than to measure absolute connectivity.

\subsection{Confirming Response Injection on UDP/5353}
\label{ssec:injection}
Active probing from five vantage points shows that the port-5353 hosts reported by Censys are injected responses rather than genuine devices. Unicast mDNS/DNS-SD service-enumeration queries to all 256 addresses in each block elicited UDP/5353 responses from ${\approx}93\%$ of addresses at four vantage points and 67\% in New York; the lower rate reflected lost queries rather than ICMP port-unreachable responses. In contrast, genuine TCP services on the same addresses responded sparsely (HTTP ${\approx}2\%$, SSH ${\leq}0.5\%$, with valid banners). Moreover, 31--37 of 39 network and broadcast (.0/.255) addresses, which cannot host individual devices, responded at every vantage point, as did baseline-absent and Censys-unlisted addresses, at the same rate as listed ones. Every response was byte-for-byte identical: a fixed 46-byte DNS response echoing the query with zero records. A control query for a random nonexistent name elicited no responses, often instead producing ICMP port-unreachable messages, indicating that the responding element selectively recognizes the well-known DNS-SD query and synthesizes an empty DNS response that scanners interpret as a live mDNS host. An earlier TCP/5353 scan elicited no responses despite the same addresses responding on TCP/80 and TCP/22, further confirming that the phenomenon is UDP-specific. Responses from impossible addresses, uniformity across blocks, transport specificity, query selectivity, and cross-continent consistency together demonstrate that the port-5353 population (${\approx}89\%$ of all plateau hosts) consists of injected middlebox responses.

\section{Conclusions}
We investigated Iran's two 2026 Internet shutdowns and their restorations across three measurement planes. Enforcement in 2022 and 2026 operates below the control plane, marking a shift from the partial BGP withdrawal observed in 2019. Restoration appears as an abrupt, centrally coordinated transition in forwarding-plane reachability with no corresponding change in BGP. Passive scanning likewise misrepresents connectivity: host visibility overshoots to approximately $3.6\times$ their pre-shutdown baseline after both restorations and rises sevenfold mid-shutdown while forwarding-plane reachability remains unchanged. Active probing shows most of these apparent hosts are injected UDP/5353 responses, including responses from network and broadcast addresses that cannot be assigned to hosts, rather than evidence of genuine connectivity. Scan visibility can also overstate in-country survival when apparent survivors are routed through foreign upstreams and never traverse the enforcement point. Each plane therefore captures a different aspect of connectivity and introduces distinct measurement biases, making cross-plane evidence essential for characterizing both shutdowns and their restorations.
\balance
\bibliographystyle{ACM-Reference-Format}
\bibliography{Bibliography}

\end{document}